\documentclass[preprint]{elsarticle}

\pdfoutput=1
\usepackage{times}
\normalfont
\usepackage{latexsym}
\usepackage[T1]{fontenc}
\usepackage[noend]{algpseudocode}
\usepackage{algorithm}
\usepackage{graphicx}
\usepackage{wrapfig}
\usepackage{verbatim}
\usepackage{amsmath}
\usepackage{stmaryrd}
\usepackage{graphics}
\usepackage{color}

\newcommand{\rdftype}{\text{rdf:type}} 
 
\newcommand{\rdfsubclpl}{\text{rdfs:subClassOf+}} 
 
\newcommand{\rdfsubproppl}{\text{rdfs:subPropertyOf+}} 
\newcommand{\rdfsdomain}{\text{rdfs:domain}} 
\newcommand{\rdfsrange}{\text{rdfs:range}} 
\newcommand{\finbox}{\phantom{.}\hfill\Box}

\newcommand{\memo}[1]{}

\newtheorem{definition}{Definition}[section]

\newtheorem{lemma}{Lemma}[section]

\newtheorem{example}{Example}[section]
\newtheorem{proof}{Proof}[section]

\begin{document}

\begin{frontmatter}

\title{Statistics of Knowledge Graphs Based On The Conceptual Schema}

\author[FAMNIT]{Iztok Savnik\corref{cor1}}
\ead{iztok.savnik@upr.si}

\author[YAHOO]{Kiyoshi Nitta}
\ead{knitta@yahoo-corp-jp}

\author[FIS,FAMNIT]{Riste \v Skrekovski}
\ead{skrekovski@gmail.com}

\author[SBG]{Nikolaus Augsten} 
\ead{nikolaus.augsten@sbg.ac.at}

\address[FAMNIT]{Faculty of mathematics, natural sciences and information technologies, University of Primorska, Slovenia}

\address[YAHOO]{Yahoo Japan Corporation, Tokyo, Japan}

\address[FIS]{Faculty of Information Studies, Novo Mesto, Slovenia} 

\address[SBG]{Department of Computer Sciences, University of Salzburg, Austria}

\cortext[cor1]{Corresponding author}

\begin{abstract}
In this paper, we propose a new approach for the computation of the statistics of knowledge graphs. We introduce a schema graph that represents the main framework for the computation of the statistics. The core of the procedure is an algorithm that determines the sub-graph of the schema graph affected by the insertion of one triple into the triple-store. We first present the algorithms that use the minimal schema and the complete schema of a knowledge graph. Furthermore, we propose an algorithm in which the size of the schema graph can be controlled---it is based on the $n$ levels from the upper part of the schema graph. We evaluate the algorithms empirically by using the Yago knowledge graph.
\end{abstract}


\end{frontmatter}

\section{Introduction}

We suppose that the basic data model of \emph{knowledge graphs} \cite{Ehrlinger2016,faerber2017ldqodfoway} is the Resource Description Framework \cite{rdf} (abbr.\ RDF). The conceptual representation of the modeled environment is in knowledge graphs provided by RDF Schema \cite{rdfschema}, which turns RDF into the knowledge representation language. Furthermore, Web Ontology Language (abbr.\ OWL) \cite{owl} extends knowledge graphs with the means to model logical relationships among the concepts and predicates, as well as the logical axioms and rules. The addition of the logical levels turns a knowledge graph into a knowledge base \cite{Brachman2004KnowledgeRA,Baader2002}.

In this paper, we present a new approach to the computation of the statistics of knowledge graphs. The proposed method is based on the conceptual schema of the knowledge graph. In this, we follow the direction established in the research area of database systems. The conceptual schema of a knowledge graph is defined solely on the basis of the languages RDF and RDF-Schema. We do not use the logical levels of knowledge graphs for the computation of the statistics. 

The \emph{conceptual schema} of a knowledge graph is defined implicitly as a part of the knowledge graph itself---it is stored in the same manner as the ordinary data. It is a graph, referred to as \emph{schema graph}, that includes class nodes related by means of the sub-class relationships and the edges linking the domain and range classes of the predicates. As we present in the paper, the nodes and edges of the schema graph represent the \emph{types} of the concrete objects and triples from the knowledge graph. 

The statistics of knowledge graphs is an essential tool used in the processes of the evaluation and optimization of queries  \cite{Neumann:2010:RES:1731351.1731354,Stocker:Etal:2008:WWW08,c-sets,gubichev14}. They are used to estimate the size of a query result and the time needed for the evaluation of a given query. The estimations are utilized when we search for the most effective query evaluation plan of a given input query. Furthermore, the statistics of large knowledge graphs are used to solve problems closely related to query evaluation. For example, they can be employed in algorithms for the partitioning of large knowledge graphs \cite{Savnik2019}. 

\subsection{Problem definition}

Let us first consider the role of statistics in database systems, in general. Suppose we have a data model, a query language, and a concrete database. The statistics of a database is used as the means to solve the problems related to query evaluation. Statistics have to be useful, in particular, to measure the identifiable and well-defined subsets of a database addressed during query processing, or some activity related to query processing. 

There are two fundamental problems related to the statistics. The first problem is to identify distinct subsets of a database that can be addressed by the queries and the relationships among them. The structure of a database obtained in this way defines the framework to which the statistics are attached. The second problem is to find an efficient algorithm to compute the statistics of the identified subsets of a database. Let us have a look at these two problems from two concrete examples, namely 1) the relational data model, and 2) the pure RDF data model without additional dictionaries. 

In the relational data model, including the query language SQL, we have a flat structure of databases. The data is separated into tables that have semantic relationships defined using the primary and foreign keys \cite{codd-rmodel}. The subsets of a database that are addressed by a given SQL query are tables stated in the FROM part of the SQL statement. The constraints that are specified in the WHERE part of the SQL statement are based on the values of attributes. Therefore, the framework to which the statistics are attached comprises the tables and the attributes of tables. The statistics of the relational database system usually include the number of rows of a relation; the number of all as well as different keys of an index; and the number of all and different values of the attributes of a table. The statistics of the index keys and attributes are often represented using the histograms to handle skew. The semantic information presented in the form of the primary and foreign keys of tables is used for the computation of the size of the joins. The framework of statistics is explicit in the case of the relational model since the framework consists of objects that are defined by a user.

The RDF data model and the query language SPARQL form the basis of recent triple-store systems \cite{Neumann:2010:RES:1731351.1731354,Gurajada:2014:TDS:2588555.2610511,jena-tdb,Harth:2007,4store,virtuoso,Zeng:2013}. These systems rely on a simple graph representation of data. The implementation of the triple-store systems is usually based on a variant of SPO indexes. The SPO indexes can be implemented as 6 SPO permutation indexes; a subset of 6 permutation indexes, most often including three permutation indexes SPO, POS, and OSP; or a subset of 6 indexes where the keys are the subsets of $\{$S,P,O$\}$. The literals that appear in triple-patterns organize the query space of triple-patterns based on the keys of SPO indexes. The set of literals form a key for the appropriate SPO index that is used to access the values of the variables from a triple-pattern. The framework for the statistics, therefore, comprises the keys of the SPO indexes. For each SPO index key, the statistics are stored in a unique aggregate index in RDF-3X \cite{Neumann:2010:RES:1731351.1731354}. The histograms are adapted from the relational systems to handle the skew. A similar approach has been taken in other triple-store systems, such as Virtuoso \cite{virtuoso} and TriAD \cite{Gurajada:2014:TDS:2588555.2610511}. 

The problem that we address in this paper is the definition of the statistics of knowledge graphs. A knowledge graph is a representation model based on the RDF and RDF-Schema data models and, including the SPARQL query language.  
A knowledge graph must include a complete conceptual schema. The classes and predicates are formally defined and then inserted into a classification hierarchy and interlinked to define the domain and range classes for each of the predicates. 
\begin{itemize}
\item The first part of the problem is the definition of a conceptual framework to which the statistics are attached. The conceptual framework has to follow the structure of the space of queries. Namely, the statistics must be able to estimate the size of triple-patterns that constitute queries. This problem and the proposed solution are introduced in Section \ref{intro-qspace}.
\item
The second part of the problem is to design an efficient algorithm for computing the statistics of a given knowledge graph. The proposed algorithm for the computation of statistics is described in Section \ref{intro-salg}.
\end{itemize}

\subsection{The proposed approach}

Let us first introduce some concepts that we use in the sequel. A \emph{schema triple} stands for a \emph{type} of triples. For instance, the schema triple (person,wasBornIn,location) is the type of all triples that include: an instance of the class person in the subject part, the predicate wasBornIn, and the instance of the class location as the object part. A \emph{schema graph} is defined by a set of classes that stand for nodes of the schema graph, a set of triples that define the classification hierarchy of classes and predicates and the set of schema triples that represent the types of triples. The \emph{stored schema graph} is a schema graph defined using the schema triples and the triples for the definition of the ontology that are \emph{stored} in a knowledge graph. We suppose that a knowledge graph includes a complete stored schema graph, which is true for most knowledge graphs \cite{Dong:2014:KVW:2623330.2623623,Hoffart201328,faerber2017ldqodfoway}.

In this paper, we propose the computation of the statistics based on the conceptual schema of a knowledge graph. We develop an index that allows answering statistical queries. The keys of the \emph{statistical index} are the schema triples that are included in a schema graph. The values of the index represent the statistical information about the schema triples (i.e., the keys). The stored statistics that we currently use are the counters of all and distinct instances of the schema triples. 

Traditionally, the computation of the statistics of a database is a problem closely related to the selectivity estimation. However, due to the complexity of the query space and, consequently, the complexity of the framework to be used as the basis for the computation of knowledge graph statistics, we address in this paper solely the problem of computing the statistics of knowledge graphs. We do not tackle the problem of the selectivity estimation except in some illustrative examples.

\subsubsection{Working example}

\begin{figure}
\begin{center}
    \includegraphics[scale=.9,bb=108 470 500 730]{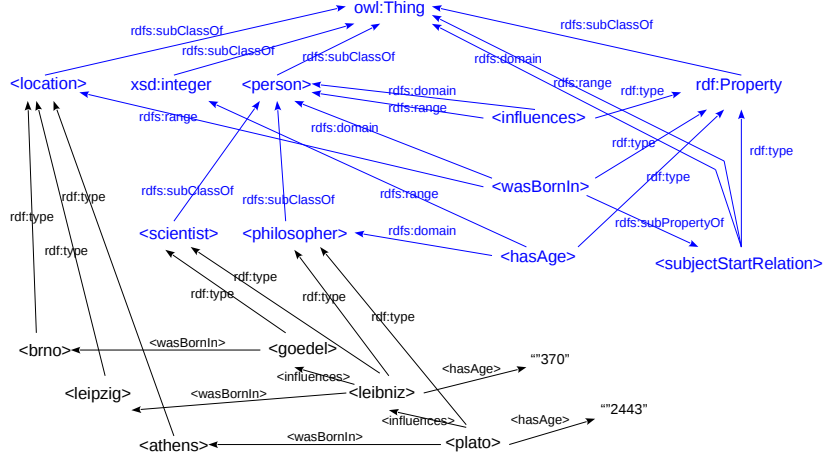}
\end{center}
\vspace{-2cm}
\caption{Triple-store \emph{Simple}.}
\label{simple}
\end{figure}

To illuminate the concepts used in the computation of the knowledge graph statistics, we use a simple knowledge graph presented in Figure \ref{simple}. The knowledge graph, referred to in the rest of the text as \emph{Simple}, includes 33 triples. Four user-defined classes are included. The class person has sub-classes scientist and philosopher. The class location represents the concept of a physical location. The schema part of \emph{Simple} is colored blue, and the data part is colored black.

\begin{example}
The stored schema graph of \emph{Simple} is represented by the schema triples (person,wasBornIn, location), (person,influence,person), (phi\-losopher,hasAge,integer) and (owl:Thing,sub\-ject\-Start\-Relation,owl:Thing), that are defined by using the predicates rdfs:domain and rdfs:range. Further, the stored schema includes the specialization hierarchy of classes and predicates---the specialization hierarchy is defined with the triples that include the predicates rdfs:subClassOf, rdfs:subPropertyOf and rdf:type. For example, the hierarchy of a class  scientist is represented by triples (scientist,\-rdfs:\-subClass\-Of,person) and (person,rdfs:subClassOf,owl:Thing).

The interpretation of a schema triple $t$ comprises a set of triples composed of individual identifiers that belong to the classes specified by the schema triple~$t$. The interpretation of the schema triple (person,wasBornIn,location), for example, are the ground triples (goedel,wasBornIn,brno), (leibniz,wasBornIn,leipzig) and (plato,was\-BornIn,\-athens). Therefore, the index key (person,wasBornIn,location) is mapped to the statistical information describing the presented instances. 
\end{example}

\subsubsection{The structure of the space of queries\label{intro-qspace}}

The structure of the space of queries can be captured by identifying the subsets of a knowledge graph that can be addressed by queries. We refer to a subset of a knowledge graph as the \emph{area} of a knowledge graph. In the relational systems, for instance, the main areas of a database system that can be addressed by queries are the tables and views stored in a database. For these areas, therefore, the statistical data is gathered.

In the case of knowledge graphs, the areas that can be addressed by the SPARQL queries have a more complex structure than areas in the relational databases. Areas of knowledge graphs correspond to the types of triples that are referred to as \emph{schema triples}. We show in the paper that the types of triples are well structured and form a partially ordered set. Consequently, the interpretations of types, i.e., the areas of a knowledge graph, also form a dual partially ordered set based on the subsumption relationship. The interpretation of the most general type of triples subsumes all the interpretations of more specific types. In the case that a triple type $t_a$ is more specific than a triple type $t_b$ than also the interpretation of $t_a$ is subsumed by the interpretation of $t_b$.
 
\emph{Triple-patterns} represent basic access methods of triple-stores. The area of a knowledge graph addressed by a triple pattern $t_p$ corresponds to the interpretation of the type of $t_p$. From this point of view, the properties of types of triples (schema triples) mentioned above make them the appropriate framework for the computation of the knowledge graph statistics. The areas of a knowledge graph that correspond to the types of triple-patterns are the areas that are the targets of SPARQL queries. Therefore, it makes sense to store the statistics for the areas defined by the \emph{stored schema triples}, the types of triples stored in a knowledge graph in the form of RDF-Schema triples. Moreover, apart from the statistics of the stored types of triples, we also selectively compute the statistics for the more specific and more general types. As presented by the following example, this allows a more precise estimation of the size of a triple-pattern. 

\begin{example}\label{example1}
The triple-pattern (?x,was\-BornIn,?y) has the type (person,was\-BornIn,\-location). Therefore, solely the instances of (person,wasBornIn,location) are addressed by the triple-pattern. Further, a join can be defined by triple-patterns (?x,was\-BornIn,?y) and (?x,hasAge,?z). The types of triple-patterns are (person,wasBornIn,\-location) and (philosopher,has\-Age,integer), respectively. We see that the variable ?x has the type person as well as philosopher. Since the interpretation of a person subsumes the interpretation of philosopher, we can safely use the schema triple (philosopher,wasBornIn,location) as the type of (?x,wasBornIn,?y).
\end{example}

From the above example, we can see that it is beneficial to store the statistics not only for the schema triples that are defined as the stored schema graph but also for more specific schema triples that include more specific classes in place of subjects and objects. The problem with storing the statistics for the set of all possible schema triples is the size of such set. Indeed, knowledge graphs include a large number of classes as well as predicates \cite{faerber2017ldqodfoway}. To be able to control the size of the schema graph, we use the stored schema graph as the reference point. The schema graph used as the skeleton for storing the statistics of a knowledge graph includes the schema triples that are up to a given number of levels more general, or, more specific, than the schema triples that form the stored schema graph.

\subsubsection{Algorithm for the computation of statistics\label{intro-salg}}

The statistics of a knowledge graph is computed by first deriving the possible types for each particular triple $t$ from the knowledge graph. The statistics are updated in the index for each of the computed types of $t$. 

Since the number of all possible types of a given triple $t$ can be substantial, the computation time for an algorithm that would compute the statistics for all possible types is unacceptably high. On the other hand, it is straightforward to compute solely the ``stored'' types of a given ground triple $t$, which means that the statistics for the stored schema of a knowledge graph can be computed efficiently. However, this algorithm does not compute the statistics for the types that are either more specific or more general than the stored types of triples. 

We propose an algorithm that selects only those types of triples that can be useful for the estimation of the size of some triple-pattern. 
The algorithm uses the types that are taken either $u$ levels above or, $l$ levels below the stored schema graph with regards to the partial ordering relationship more-general/more-specific defined on the schema triples.

\subsection{Contributions}

The main contribution of the paper is a new approach to the definition of the statistics of knowledge graphs. The schema graph is proposed to be used as a skeleton for the computation of statistics. The statistics are computed for each of the schema triples that comprise the stored schema graph. Furthermore, we compute the statistics for the selected schema triples that are more general or more specific than some schema triples from the stored schema graph. In this way, we can obtain more detailed statistics that allow more precise selectivity estimation of triple-patterns. To the best of our knowledge, this is the first proposal to use schema triples, i.e., the types of triples, as the basis for the computation of the knowledge graph statistics.

The types of triples are, in many respects, similar to the relational schemata. They are general in the sense that similar solutions to those proposed in the relational systems apply to the schema triples. For instance, the histograms can be used to handle the non-uniform distribution of the predicates. Furthermore, the schema triples can be compared to the characteristics sets, as defined in \cite{c-sets,gubichev14}. The statistics can be computed for the pairs of connected schema triples to capture correlations among predicates in the estimation of the size of joins. 

The second original contribution of the presented work is the proposal of an efficient algorithm for the computation of the statistics of a knowledge graph. The algorithm selects the schema graph that includes the schema triples from the stored schema graph as well as the schema triples that are $u$ levels more general, or $l$ levels more specific than the schema triples from the stored schema graph. The parameters $u$ and $l$ control the number of the schema triples for which the statistics are computed. Therefore, the parameters $u$ and $l$ determine the amount of the computation needed to compute the statistics of a knowledge graph.

\subsection{Organization}

The paper is organized as follows. Section 2 presents a formal view of a knowledge graph. It introduces the concepts to be used for the definition of the framework and method for computing the statistics of a knowledge graph. The denotational semantics of the schema triples and the schema graphs is defined.

Section 3 describes three algorithms for the computation of the statistics. The first algorithm, presented in Section \ref{1st-stat-proc}, computes the statistics for the stored schema graph. The second algorithm, given in Section \ref{2nd-stat-proc}, computes the statistics of all possible schema triples. From these two algorithms, we construct the third algorithm, in Section \ref{3rd-stat-proc}, that combines the features of the first two and restricts the final schema graph by taking into account only the schema triples some levels below and some levels above the schema triples from the stored schema graph.

Section 4 introduces the concept of a key, which is used to count the instances of a given schema triple. Section \ref{keys-types} defines the key, the type of a key, the underlined type of a key, and the complete type of a key, formally. In Section \ref{comp-stat-keys}, we show the different ways we can count the keys. Firstly, we can count either all or just the distinct keys of schema triples. Secondly, we discern between the bound and the unbound ways of counting the instances of a given schema triple. The counting algorithm is either strictly bound to the schema graph, or, does not take into account the schema graph, as presented in Section \ref{cnt-bound-unbound}.

The experimental evaluation of the algorithm for the computation of statistics is given in Section 5. We first present the computation of the statistics on the running example introduced in Figure \ref{simple}. Secondly, Section \ref{exp-yago} presents a series of experiments on the Yago knowledge graph. 

Finally, the related work is presented in Section 6, and the concluding remarks, together with some directions of our further work, are given in Section 7.

\section{Conceptual schema of a knowledge graph\label{ts-schema}}

The formal definition of the schema of a knowledge graph is based on the RDF \cite{rdf} and RDF-Schema \cite{rdfschema} data models. Let $I$ be the set of URI-s, $B$ be the set of blanks and $L$ be the set of literals. Let us also define sets $S = I\cup\/B$, $P = I$, and $O = I\cup\/B\cup\/L$.

A \emph{RDF triple} is a triple $(s,p,o)\in\/S\times\/P\times\/O$. An \emph{RDF graph} $g\subseteq\/S\times\/P\times\/O$ is a set of triples. The set of all graphs is denoted as $G$. We state that an RDF graph $g_1$ is a \emph{sub-graph} of $g_2$, denoted as $g_1 \subseteq g_2$, if all triples of $g_1$ are also triples of $g_2$. We call the elements of the set, $I$, \emph{identifiers} to abstract away the details of the RDF data model \cite{b3s-graph-algebra}. 

The predicates that are used for the definition of a conceptual schema of a knowledge graph are rdf:type, rdfs:subClassOf, rdfs:subPropertyOf, rdfs:Domain and rdfs:\-Range. We assume that the conceptual schema of a knowledge graph is defined. 

\subsection{Identifiers\label{tsform-identifiers}}

We define a set of concepts to be used for a more precise characterization of identifiers $I$. \emph{Individual identifiers}, denoted as set $I_i$, are identifiers that have specified their classes using property rdf:type. \emph{Class identifiers} denoted as set $I_c$, are identifiers that are sub-classes of the top class $\top$ of ontology---Yago, for instance, uses the top class owl:Thing. 

The complete set of identifiers $I$ is, therefore, composed of the individual and class identifiers, or, $I=I_i\cup\/I_c$. The individual identifiers can be further divided into individual identifiers that stand for the objects and things, and, particular types of individual identifiers that stand for predicates. 

\emph{Predicate identifiers} denoted as set $I_p$, are identifiers that represent RDF predicates. The predicate identifiers are, from one perspective, very similar to the class identifiers. Indeed the predicates can have sub-predicates in the same manner as the classes have sub-classes. However, while classes have instances, predicates do not have instances. Predicates $I_p$ are individual identifiers that are the instances of rdf:Property. 

The interpretation of a given class identifier $c$ is a set of individual identifiers that are the instances of $c$. The interpretation of $c$ is denoted $\llbracket\/c\rrbracket_g$. The interpretation of a class includes the interpretations of all its sub-classes. Let us now formally define the interpretation of classes. 

\begin{definition}
  Let $g\in\/G$ be a graph and let $c$ be a class identifier that is
  a part of the graph $g$. The interpretation of $c$ is
  \begin{displaymath}
  \begin{array}{l}
     \llbracket c\rrbracket_g=\{i\mid (i,\rdftype,c)\in\/g\ \lor \\
     \phantom{\llbracket c\rrbracket_g=\{i\mid\ }
                       \exists\/c'((c',\rdfsubclpl,c)\in\/g\ \land \\
     \phantom{\llbracket c\rrbracket_g=\{i\mid\ \exists\/c'(}
                       (i,\rdftype,c')\in\/g)\},
  \end{array}
  \end{displaymath}
  where ``+'' denotes the direct or transitive use of a given property. $\finbox$
\end{definition}

Note that $\llbracket\/c\rrbracket_g\subseteq\/I_i$. 

We define the partial ordering of identifiers by using the relationships rdf:type, rdfs:subClassOf and rdfs:subPropertyOf. The top of the partial ordering is composed of the most general classes together with the top-class $\top$. The bottom of the partial ordering of identifiers includes the individual identifiers from $I_i$. 

\begin{definition}
  Let $g\in\/G$ and $i_1,i_2\in\/I$. Relationship
  \emph{is-more-specific} $i_1\preceq\/i_2$ is defined in the
  following way with respect to the classification of identifiers
  $i_1$ and $i_2$:
  \begin{enumerate}
  \item if $i_1,i_2\in\/I_i$ then $i_1=i_2$;
  \item if $i_1\in\/I_i, i_2\in\/I_c$ then $i_1\in\llbracket\/i_2\rrbracket_g$;
  \item if $i_1,i_2\in\/I_c$ then $i_1=i_2\lor\/(i_1,\rdfsubclpl,i_2)\in\/g$; 
  \item if $i_1,i_2\in\/I_p$ then $i_1=i_2\lor\/(i_1,\rdfsubproppl,i_2)\in\/g$.
  \end{enumerate}
\end{definition}

Relationship ``$\preceq$'' is reflexive, transitive, and anti-symmetric, i.e., it defines a partial order relationship.

Let us now define another interpretation of identifiers called a \emph{natural interpretation}. This interpretation maps identifiers $i$ to the sets of all identifiers that are more specific than $i$. The individual identifiers are mapped to themselves, while the class identifiers are mapped to the sets of all identifiers that are more specific. The denotational semantics of natural interpretation is defined as follows. 

\begin{definition}
  Let $g\in\/G$ and $i\in\/I$. The \emph{natural interpretation} of $i$ is
  $$\llbracket\/i\rrbracket^*_g = \{i'| i'\in\/I \land i'\preceq\/i\}.$$
\end{definition}

The partial ordering relationship $\preceq$ and the semantic functions $\llbracket\rrbracket_g$ and $\llbracket\rrbracket_g^*$ are consistent. The relationship $\preceq$ between two identifiers implies the semantic subsumption of the ordinary interpretations as well as natural interpretations of the given identifiers. Here we show solely that $\preceq$ implies subsumption of natural interpretations. 

\begin{lemma}
  \label{lemma1}
  Let $g\in\/G$ be a graph and $i_1,i_2\in\/I$ be identifiers. If the relationship $i_1\preceq\/i_2$ holds, then  $\llbracket\/i_1\rrbracket^*_g\subseteq\llbracket\/i_2\rrbracket_g^*$. 
\end{lemma}

\begin{proof}
Suppose now $i_1\preceq\/i_2$. For every $x\in\llbracket\/i_1\rrbracket^*_g$ it holds $x\preceq\/i_1$. By transitivity, $x\preceq\/i_2$ holds. This implies $\llbracket\/i_1\rrbracket^*_g\subseteq\llbracket\/i_2\rrbracket_g^*$. 
\end{proof}

\subsection{Triples and graphs\label{tsform-triples}}

We differentiate between the ground, abstract, and schema triples. The \emph{ground triples} can only include individual identifiers as well as literals in the place of a third component. The \emph{abstract triples} include at least one class identifier. A triple that includes the classes and predicates solely is called the \emph{schema triple}. For example, the triple (plato,was\-Born\-In,athens) is a ground triple, and its type, the triple (person,was\-Born\-In,location), is a schema triple. The schema triples are defined more precisely in Definition \ref{schematriple}.

We extend the set of identifiers with the top class owl:Thing. We assume in the sequel that owl:Thing $\in\/I$, and for every $i\in\/I$, it holds that $i\preceq$ owl:Thing. 

The partial ordering relation $\preceq$ is extended to triples. A triple $t_1$ is more specific or equal to a triple $t_2$ if all components of $t_1$ are more specific than the components of $t_2$. 

\begin{definition}
  \label{tripleposet}
  Let $t_1=(s_1,p_1,o_1)$ and $t_2=(s_2,p_2,o_2)$ such that
  $t_1,t_2\in\/S\times\/P\times\/O$. Triple $t_1$ is more specific or
  equal to triple $t_2$, denoted as
  $t_1\preceq\/t_2\iff\/s_1\preceq\/s_2$, $p_1\preceq\/p_2$ and
  $o_1\preceq\/o_2$.
\end{definition}
 
Let us now define a schema triple that represents a \emph{type} of a set of ground triples. A schema triple is either a stored schema triple or a schema triple that is related to some stored schema triple using the relationship $\preceq$ (more general/specific or equal). 

A stored schema triple $(s,p,o)$ is a triple that includes a predicate $p$, a subject $s$ that is a domain class of $p$, and an object $o$ that is a range class of $p$. Formally, a stored schema triple is defined as follows. 

\begin{definition}
Let $g\in\/G$ be a graph. A stored schema triple is a triple $t=(s,p,o)$ such that $(p,\rdftype,\text{rdf:Property})\in\/g$, $(p,\rdfsdomain,s)\in\/g$ and \\ $(p,\rdfsrange,o)\in\/g$. 
\end{definition}

The schema triple can now be defined as follows. 

\begin{definition}
\label{schematriple}
Let $g\in\/G$ be a graph and let \emph{sst} be a set of all stored schema triples of $g$. A schema triple is a triple $t$ such that either $t\in\text{sst}$, or there exists a stored schema triple $t_s\in\/\text{sst}$ and $t\preceq\/t_s\lor\/t\succeq\/t_s$.
\end{definition}

The above definition ties the schema triples to the partially ordered set of triples introduced in Definition \ref{tripleposet}. Moreover, a  schema triple must be either more general or, more specific than some stored schema triple, and therefore, it must represent a legal type in a given knowledge graph. Other triples that include the classes and predicates solely are not legal types in a given knowledge graph.

\begin{example}
Let us present examples of stored schema triples and the schema triples that are more specific or more general then stored schema triples. The example is based on the knowledge graph \emph{Simple} presented in Figure \ref{simple}. The stored schema triples in \emph{Simple} include (person,wasBornIn,location), (person,influences,person) and (philosopher,hasAge,integer). Note that the stored schema triples are represented utilizing the predicates rdfs:domain and rdfs:range. 

The schema triples that are candidates for the keys of a statistical index include (scientist,\-wasBornIn,location), (philosopher,wasBornIn,location), (scientist,influ\-en\-ces,\-sci\-entist), and so on. The triple (scientist,wasBornIn,location) is a schema triple because of the stored schema triple (person,wasBornIn,location) where the class person is a super-class of the class scientist. $\finbox$
\end{example}

We can now define the interpretation of a schema triple, similarly to the case of identifiers. The interpretation function $\llbracket\/t\rrbracket_g$ maps a schema triple $t$ to the set of ground triples that are the instances of a given schema triple $t$. The natural interpretation function $\llbracket\/t\rrbracket_g^*$ maps a schema triple $t$ to the set of triples that are more specific than $t$, or equal to $t$.

\begin{definition}
  Let $t=(s,p,o)$ be triples such that
  $(s,p,o)\in\/S\times\/P\times\/O$. The interpretation functions
  $\llbracket\rrbracket_g$ and $\llbracket\rrbracket_g^*$ are defined as:
  \begin{itemize}
  \item $\llbracket\/t\rrbracket_g=\{(s',p',o')| (s',p',o')\in\/g\land\/s'\in\llbracket\/s\rrbracket_g \land 
                              p'\in\llbracket\/p\rrbracket_g \land o'\in\llbracket\/o\rrbracket_g \}$; 
  \item $\llbracket\/t\rrbracket^*_g=\{(s',p',o')| (s',p',o')\in\/g\land\/s'\preceq\/s \land 
                              p'\preceq\/p \land o'\preceq\/o \}$. $\finbox$
  \end{itemize}
\end{definition}

The consistency of the partially ordered set of triples with the subsumption hierarchy of the interpretations of the schema triples can be shown similarly as in Lemma~\ref{lemma1}. The semantic relationship $\preceq$ implies the subsumption among the interpretations of schema triples. For this reason, the space of triples from a given knowledge graph is ordered into the subsumption hierarchy that reflects the partial ordering of triple types. 

The set of all stored schema triples together with the triples that define the classification hierarchy of the classes and predicates form the \emph{stored schema graph}. The stored schema graph of a given knowledge graph $g$ is formally defined as follows. 

\begin{definition}
  Let $g\in\/G$ be a graph. The stored schema graph $ssg$ is defined as
  \begin{displaymath}
  \begin{array}{ll}
     ssg = \{(s,p,o)|(p,\rdftype,\text{rdf:Property})\in\/g\ \land \\
     \phantom{ssg = \{} (p,\rdfsdomain,s)\in\/g\land(p,\rdfsrange,o)\in\/g\}\ \cup \\
     \phantom{ssg =\ } \{(s,p,o)|(s,p,o)\in\/g\land\/((s,o\in\/I_c\land p=\text{rdfs:subClassOf})\ \lor \\
     \phantom{ssg = \{} (s,o\in\/I_p \land p=\text{rdfs:subPropertyOf}))\}. \\
  \end{array}
  \end{displaymath}
\end{definition}

The stored schema graph defines the structure of ground triples in $g$. Each ground triple is an instance of at least one schema triple $t_s\in\/ssg$. 

\begin{example}
The stored schema of the knowledge graph \emph{Simple} is colored blue in Figure~\ref{simple}. The stored schema graph is derived from the graph by replacing the triples using predicates rdfs:domain and rdfs:range with the schema triples. The schema triples included in \emph{Simple} are  (person,wasBornIn,location), (person,influen\-ces,person), (phi\-losopher,hasAge,integer) and (owl:Thing,\-subjectStart\-Relation,\-owl:\-Thing). 

Note that the meta-schema predicates are not part of this simple knowledge graph. For this reason, we suppose that the meta-schema triples have the subject and object components set to owl:Thing. The schema triple for the predicate rdf:type, for instance, is (owl:Thing,rdf:type,owl:Thing). $\finbox$
\end{example}

Let us now define the concept of a \emph{schema graph} formally. A schema graph of $g$ includes the stored schema graph of $g$ as well as a set of schema triples that are either more general or more specific than the stored schema triples. 

\begin{definition}
  Let $g\in\/G$ be a graph and $ssg$ be a stored schema graph of $g$. A complete schema graph $csg$ is defined as $$csg=ssg\cup\{t|t_s\in\/ssg\land(t\preceq\/t_s\lor\/t_s\preceq\/t)\}.$$ A schema graph $sg$ is a graph such that $ssg\subseteq\/sg\subseteq\/csg$.
\end{definition}

Every schema graph of $g$ includes at least the stored schema graph. A schema graph includes additional schema triples that are used to capture the statistics of possible types of triple-patterns. 

\subsection{Triple-patterns\label{3store-tp-form}}

We assumed the existence of a set of variables $V$. The following definition introduces triple-patterns as triples that can include variables.

\begin{definition}
A \emph{triple pattern}
$(s,p,o)\in(S\cup\/V)\times\/(P\cup\/V)\times\/(O\cup\/V)$ is a triple
that includes at least one variable.
\end{definition}

For the presentation of the semantics of triple-patterns, we define an alternative way of accessing the components of triples. The components of triple $t=(s,p,o)$ can be accessed in, similarly to the elements in an array, as $t[1]=s$, $t[2]=p$ and $t[3]=o$.

Let $tp=(s,p,o)$ be a triple pattern and the set $tp_v\subseteq\/\{1,2,3\}$ be a set of indices of components that are variables, i.e., $\forall\/j\in\/tp_v: tp[j]\in\/V$. The interpretation of a triple pattern $tp$ is the set of triples $t\in\/g$ such that $t$ includes any value in place of variables indexed by the elements of $tp_v$, and, has the values of other components equal to the corresponding $tp$ components. 

\begin{definition}
Let $g\in\/G$ be a graph, $tp$ a triple-pattern, and, $tp_v$ a set of indices identifying variables of $tp$. The interpretation of $tp$ is:
  \begin{displaymath}
  \llbracket\/tp\rrbracket_g=\{t\mid\/t\in\/g\land
  \forall j\in\{1,2,3\}\setminus\/tp_v: t[j]=tp[j]\}
  \end{displaymath}
\end{definition}

The type of a triple-pattern is a schema triple such that the interpretation of the schema triple subsumes the interpretation of the triple-pattern.

\begin{definition}
Let $tp=(s,p,o)$ be a triple-pattern and $t_{tp}=(t_s,t_p,t_o)$ be a schema triple, such that $t_s,t_o\in\/I_c$ are classes and $t_p\in\/I_p$ is the property. The schema triple $t_{tp}$ is the type of $tp$ if and only if
  \begin{displaymath}
    \llbracket\/tp\rrbracket_g\subseteq\llbracket\/t_{tp}\rrbracket^*_g.
  \end{displaymath}
\end{definition}

Note that we only specify the semantic conditions for the definition of the type of triple-patterns. The algorithm for the derivation of the type of a triple-pattern is beyond the scope of the presented research; it is defined in \cite{b3s-type-infer}.

\begin{example}
We now present a few examples of the natural interpretation function $\llbracket\rrbracket^*$, within the context of the example graph $g$ given in Figure \ref{simple}. The natural interpretation of the schema triple $t_{s_1}=$ (owl:Thing,rdf:type,person) includes four triples. 

\begin{center}
\begin{tabular}{lll}
  $\llbracket\/t_{s_1}\rrbracket^*_g$ & $=$ & $\{$(plato,rdf:type,philosopher), (leibniz,rdf:type,philosopher),\\
  & &  \phantom{$\{$}(leibniz,rdf:type,scientist), (goedel,rdf:type,scientist)$\}$. \\
\end{tabular}
\end{center}

The natural interpretation of the schema triple $t_{s_2}=$(person,wasBornIn,location) has three triples.

\begin{center}
\begin{tabular}{lll}
$\llbracket\/t_{s_2}\rrbracket^*_g$ & $=$ & $\{$(plato,wasBornIn,athens), (leibniz,wasBornIn,leipzig), \\
                               &       & \phantom{$\{$}(goedel,wasBornIn,brno)$\}$ \\ 
\end{tabular}
\end{center}
\end{example}

\section{Computing statistics}

The statistical index is implemented as a dictionary where keys represent schema triples, and the values represent the statistical information for the given keys. For instance, the index entry for the schema triple (person,wasBornIn,\-location) represents the statistical information about the triples that have the instance of a person as the first component, the property wasBornIn as the second component and the instance of location as the third component. 


The main procedure for the computation of the statistics of a knowledge graph is presented as Algorithm \ref{statistics-ts}. The statistic is computed for each triple $t$ from a given knowledge graph. The function \textsc{statistics-triple}$(t)$ represents one of three functions: \textsc{statistics-stored}$(t)$, \textsc{statistics-all}$(t)$ and \textsc{statistics-levels}$(t)$, which are presented in detail in the sequel. Each of the functions computes, in the first phase, the set of schema triples, which include the given triple, $t$, in their natural interpretations $\llbracket\rrbracket^*$. In the second phase, the function \textsc{statistics-triple} updates the corresponding statistics for the triple $t$. 

\begin{algorithm}
\normalsize
\begin{algorithmic}[1]
\Procedure{compute-statistics}{$ts:$ knowlege-graph}
   \ForAll{$t\in\/ts$}
      \State \Call{statistics-triple}{$t$}
   \EndFor
\EndProcedure
\end{algorithmic}
\caption{Procedure \textsc{compute-statistics}($ts:$ knowledge-graph)}
\label{statistics-ts}
\end{algorithm}

This section is organized as follows. The procedure \textsc{statistics-stored} that computes the statistics for the stored schema graph is described in Section \ref{1st-stat-proc}. The second procedure \textsc{statistics-all} computes statistics for all possible types of $t$---it is presented in Section \ref{2nd-stat-proc}. The procedure \textsc{statistics-levels} selects the schema triples that are some levels more specific or more general than the schema triples from the stored schema graph. The procedure is presented in Section \ref{3rd-stat-proc}. Finally, the procedure \textsc{retrieve-stat} retrieves the statistics for a given schema triple. In the case that the schema triple is not a key of the statistical index, then the statistics are approximated from the statistics of the existent keys. This procedure is presented in Section \ref{retrieve-stat}. 

\subsection{Statistics of the stored schema graph\label{1st-stat-proc}}

The first procedure for computing the statistics is based on the schema information that is a part of the knowledge graph. As we have already stated in the introduction, we suppose that the knowledge graph would include a complete schema, i.e., the definition of classes, predicates, as well as the domains and ranges of predicates.  

\begin{algorithm}
\normalsize
\begin{algorithmic}[1]
\Function{statistics-stored}{$t=(s,p,o):$ triple-type}
   \State $g_p \gets \{p\}\cup\{c_p| (p,\text{rdfs:subPropertyOf}+,c_p) \in g\}$

   \ForAll{$p_g\in\/g_p$}
      \State $d_p \gets \{ t_s | (p_g,\text{rdfs:domain},t_s)\in\/g\}$
      \State $r_p \gets \{ t_o | (p_g,\text{rdfs:range},t_o)\in\/g\}$
      \ForAll{$t_s\in\/d_p, t_o\in\/r_p$}
          \State \Call{Update-Statistics}{($t_s,p_g,t_o),t$}
      \EndFor
   \EndFor
\EndFunction
\end{algorithmic}
\caption{Function \textsc{statistics-stored}($t:$ triple-type)}
\label{statistics-stored}
\end{algorithm}

Let us now present Algorithm \ref{statistics-stored}. We assumed that $t=(s,p,o)$ was an arbitrary triple from graph $g\in\/G$. First, the algorithm initializes set $g_p$ in the line~2 to include the element $p$, and the transitive closure of $\{p\}$ computed with respect to the relationship rdfs:subPropertyOf to obtain all more general properties of $p$. Note that '+' denotes one or more application of the predicate rdfs:subPropertyOf. 

After the set $g_p$ is computed, the domains and the ranges of each particular property $p_g\in\/g_p$ are retrieved from the graph in lines 4-5. After we have computed all the sets, including the types of $t$'s components, the schema triples can be enumerated by taking property $p_g$ and a pair of the domain and range classes of $p_g$. The statistics are updated using the procedure \textsc{update-statistics} for each of the generated schema triples in line 7. A detailed description of the procedure \textsc{update-statistics} is given in Section \ref{updt-sttstcs}.

\begin{example}
Let us now present an example of how statistics of the stored schema graph is computed. The triple used in the example is (leibniz,influences,goedel). First, we compute the transitive closure of the set $\{$influences$\}$ by means of relationship rdfs:subPropertyOf obtaining, since there are no predicates that are more general than predicate influences, the same set $g_p=\{$influences$\}$. Further, by using triples   $($influences,rdfs:domain,person$)$ and $($influences,rdfs:range,person$)$, we can compute the sets $d_p = \{$person$\}$ and $r_p = \{$person$\}$ that represent the domains and ranges of the property influences. We generate the schema triple $($person,influences,person$)$, in this way. Since the predicate influences is the only element of the set $g_p$, there is nothing more to do. Therefore, no additional schema triples are generated. Finally, the procedure \textsc{update-statistics} is called for the generated schema triple.  $\finbox$ 
\end{example}

\subsection{Statistics of all schema triples\label{2nd-stat-proc}}

Let $t=(s,p,o)$ be arbitrary triple from graph $g\in\/G$. The set of schema triples for a given triple $t$ is obtained by first computing all possible classes of the components $s$ and $o$, and all existing super-predicates of $p$. In this way, we obtain the sets of classes $g_s$ and $g_o$, and, the set of predicates $g_p$. These sets are then used to generate all possible schema triples, i.e., all possible types of $t$. The statistics are updated for the generated schema triples. 

\begin{algorithm}
\normalsize
\begin{algorithmic}[1]
\Function{statistics-all}{$t=(s,p,o):$ triple-type}
   \State $g_s \gets \{ t_s | \text{is\_class}(s) \land t_s = s \lor \neg\text{is\_class}(s) \land (s,\text{rdf:type},t_s)\in\/g\}$ 
   \State $g_o \gets \{ t_o | \text{is\_class}(o) \land t_o = o \lor \neg\text{is\_class}(o) \land (o,\text{rdf:type},t_o)\in\/g\}$ 
   \State $g_s \gets g_s\cup\{c_s'| c_s\in\/g_s \land (c_s,\text{rdfs:subClassOf}+,c_s') \in g\}$
   \State $g_p \gets \{p\}\cup\{c_p| (p,\text{rdfs:subPropertyOf}+,c_p) \in g\}$
   \State $g_o \gets g_o\cup\{c_o'| c_o\in\/g_o \land (c_o,\text{rdfs:subClassOf}+,c_o') \in g\}$
   \ForAll{$c_s\in\/g_s, c_p\in\/g_p, c_o\in\/g_o$}
         \State \Call{Update-Statistics}{$(c_s,c_p,c_o),t$}
   \EndFor
\EndFunction
\end{algorithmic}
\caption{Function \textsc{statistics-all}($t:$ triple-type)}
\label{statistics2}
\end{algorithm}

The procedure for the computation of statistics for a triple $t$ is presented in Algorithm \ref{statistics2}. The triple $t$ includes the components $s$, $p$ and $o$, i.e., $t=(s,p,o)$. The procedure \textsc{statistics-all} computes in lines 1-6 the sets of types $g_s$, $g_p$ and $g_o$ of the triple elements $s$, $p$ and $o$, respectively. The set of types $g_s$ is in line 2 initialized by the set of the types of $s$, or, by $s$ itself when $s$ is a class. The set $g_s$ is then closed by means of the relationship refs:subClassOf in line 4. Set $g_o$ is computed in the same way as set $g_s$. The set of predicates $g_p$ is computed differently. Set $g_p$ obtains the value by closing set $\{p\}$ using the relationship rdfs:subPropertyOf in line 5. Indeed, predicates play a similar role to classes in many knowledge representation systems \cite{cyc}. 

The schema triples of $t$ are enumerated in the last part of the procedure \textsc{statistics-all} by using sets $g_s, g_p$ and $g_o$ in lines 7-8. For each of the generated schema triples $(c_s,c_p,c_o)$ the interpretation $\llbracket(c_s,c_p,c_o)\rrbracket_g^*$ includes the triple $t=(s.p.o)$. Since $g_s$ and $g_o$ include all classes of $s$ and $o$, and, $g_p$ includes $p$ and all its super-predicates, all possible schema triples in which interpretation includes $t$ are enumerated. 

\begin{example}
Let us first present an example of computing sets of classes $g_s,g_p,g_o$ for a triple (plato,wasBornIn,athens). 

The set $g_s$ initially includes the class philosopher because of the triple $($plato,rdf:\-type,\-philosopher$)$. Since there are no other types of plato we can now compute the transitive closure of the initial set $g_s$ by using relationship rdfs:subClassOf. We thus obtain the final set $g_s=\{$philosopher,person,rdf:Thing$\}$.

Set $g_p$ initially includes the predicate wasBornIn. The transitive closure of $g_s$ is computed by using the relationship rdfs:subPropertyOf yielding the set $\{$wasBornIn, subjectStartRelation$\}$.

Similarly to the case of $g_s$ we compute the set $g_o=\{$location,owl:Thing$\}$.

The schema triples that are generated by Algorithm \ref{2nd-stat-proc} are: (person,wasBornIn,\-location), (person,subjectStartRelation,location), (philosopher,wasBornIn,location), \-(philosopher,subjectStartRelation,location), (owl:Thing,wasBornIn,location), \-(owl: \-Thing,subjectStartRelation,location), (person,was\-BornIn,\-owl:Thing), (person,subject\-StartRelation,owl:Thing), (philosopher,wasBornIn,owl:Thing), (philosopher,subject \-StartRelation,owl:Thing), (owl:Thing,wasBornIn,owl:Thing), and (owl:Thing,\-subject\-StartRelation,\-owl:Thing). $\finbox$
\end{example}

\subsection{Statistics of a strip around the stored schema graph\label{3rd-stat-proc}}

The procedures for the computation of the statistics presented above suffer from some practical problems. The first procedure computes statistics solely for the schema triples that are included in the knowledge graph. While these statistics can be readily used to estimate the size of the results of arbitrary triple-patterns, the estimates may not be precise. 

\begin{algorithm}
\normalsize
\begin{algorithmic}[1]
\Function{statistics-levels}{$t=(s,p,o):$ triple-type, $k:$ integer}

   \State $g_p \gets \{p\}\cup\{t_p| (p,\text{rdfs:subPropertyOf}+,t_p) \in g\}$
   \State $g_s \gets \{ t_s | \text{is\_class}(s) \land t_s = s \lor \neg\text{is\_class}(s) \land (s,\text{rdf:type},t_s)\in\/g\}$ 
   \State $g_o \gets \{ t_o | \text{is\_class}(o) \land t_o = o \lor \neg\text{is\_class}(o) \land (o,\text{rdf:type},t_o)\in\/g\}$ 
      
   \State $g_s \gets g_s\cup\{c_s'| c_s\in\/g_s \land (c_s,\text{rdfs:subClassOf}+,c_s') \in g\}$
   \State $g_o \gets g_o\cup\{c_o'| c_o\in\/g_o \land (c_o,\text{rdfs:subClassOf}+,c_o') \in g\}$
         
   \State $d_p \gets \{c_s|(p,\text{rdfs:domain},c_s)\in\/g\}$
   \State $r_p \gets \{c_s|(p,\text{rdfs:range},c_s)\in\/g\}$

   \State $s_s \gets \{c_s| c_s\in\/g_s\land \exists\/c_s'\in\/d_p((c_s\succeq\/c_s'\land\textsc{dist}(c_s,c_s')\le\/u)\ \lor$
   \State \phantom{$s_s \gets \{c_s| c_s\in\/g_s\land \exists\/c_s'\in\/d_p($}$(c_s\preceq\/c_s'\land\textsc{dist}(c_s,c_s')\le\/l))\}$
   \State $s_o \gets \{c_o| c_o\in\/g_o\land \exists\/c_o'\in\/r_p((c_o\succeq\/c_o'\land\textsc{dist}(c_o,c_o')\le\/u)\ \lor$
   \State \phantom{$s_o \gets \{c_o| c_o\in\/g_o\land \exists\/c_o'\in\/r_p($}$(c_o\preceq\/c_o'\land\textsc{dist}(c_o,c_o')\le\/l))\}$

   \ForAll{$c_s\in\/s_s, p_p\in\/g_p, c_o\in\/s_o$} 
      \State \Call{Update-Statistics}{($c_s,p_p,c_o),t$}
   \EndFor
\EndFunction
\end{algorithmic}
\caption{Function \textsc{statistics-levels}($t:$ triple-type, $k:$ integer)}
\label{statistics-levels}
\end{algorithm}

In the procedure \textsc{statistics-stored}, we did not compute the statistics of the schema triples that are either more specific or more general than the schema triples that are included in the knowledge graph for a given predicate. For instance, while we do have the statistics for the schema triple (person,wasBornIn,location) since this schema triple is part of the knowledge graph, we do not have the statistics for the schema triples (scientist,wasBornIn,location) and (philosopher,wasBornIn,location). 

The procedure \textsc{statistics-all} is in a sense the opposite of the procedure \textsc{statistics-stored}. Given the parameter triple $t$, the procedure \textsc{statistics-all} updates statistics for \emph{all} possible schema triples which interpretation includes the parameter triple $t$. The number of all possible schema triples may be much too large for a knowledge graph with a rich conceptual schema. The conceptual schema of Yago knowledge graph, for instance, includes a half-million classes. 

The stored schema triples are used as reference points in the description of the \textsc{statistics-levels}. The procedure \textsc{statistics-levels} is an extension of the first and the second procedures. It first determines all possible schema triples of a given triple $t=(s,p,o)$ by computing the classes of components $s$ and $o$, and, the super-properties of the component $p$ in the same way as in the procedure \textsc{statistics-all}. Next, similarly to the procedure \textsc{statistics-stored}, it uses the stored schema as the basis for the definition of the area composed of the schema triples that are taken into account for the computation of the statistics. The schema used for the statistics is defined as the intersection of the set of schema triples that are the types of $t$ with the set of schema triples located at some levels around the stored schema. Let us now present the procedure for the computation of statistics given in Algorithm \ref{statistics-levels} in more detail. 

The parameters of the procedure \textsc{statistics-levels} are a triple $t=(s,p,o)$ and two integer numbers $u$ and $l$ that define the area of the schema used for storing statistics. The number $u$ specifies the number of levels above the stored schema, and $l$ specifies the number of levels below the stored schema. 

The first part of the algorithm, written in lines 1-6, is precisely the same as in the procedure \textsc{statistics-all}, i.e., all possible classes of $s$ and $o$, as well as, all super-predicates of $p$ are determined.  The second part of the algorithm, given in lines 7-8, follows the procedure \textsc{statistics-stored}: it determines the domains and ranges of the predicate $p$ from the stored schema. Finally, the third part of the algorithm, given in lines 9-12, selects the schema triples $l$ levels below and $u$ levels above the stored schema from all possible schema triples that represent the types of $(s,p,o)$. 

We separately compute the classes to be used to enumerate the selected schema triples in lines 9-12. The set $s_s$ is a set of classes $c_s$ such that $c_s\in\/g_s$ and every $c_s$ is either more general or more specific than some $c\in\/d_s$. Furthermore, the distance between $c_s$ and $c$ must be less than $u$ when $c_s\succeq\/c$, or, less than $l$ when $c_s\preceq\/c$. The set $s_o$ is defined very similarly except that it includes the selected classes for the O component of the schema triples. The classes from $s_s$ and $s_o$ represent the domain and range classes of the schema triples selected for the schema of the statistics. 

Finally, the statistic is updated in lines 13-14 for all schema triples that are generated using the Cartesian product of the sets $s_s$, $g_p$, and $s_o$.  

\subsection{Retrieving statistics of a schema triple\label{retrieve-stat}}

Let us suppose that we have the statistical index of a knowledge graph computed using the algorithm \textsc{statistics-levels}, and, that the statistics are based on the schema triples from $u$ levels above the stored schema triples and $l$ levels below the stored schema. The statistical index is implemented as a key-value dictionary where the key is a schema triple, and the value represents the statistical entry for a given key. We call the set of schema triples that represent the keys of the statistical index the \emph{schema of statistics} $S$. 

When we want to retrieve the statistical data for a schema triple $t=(s,p,o)$, we have two possible scenarios. The first is that the schema triple $t$ is stored in the dictionary. The statistics are, in this case, directly retrieved from the statistical index. This is the expected scenario, i.e., we expect that the statistical index includes all schema triples that can represent the types of triple-patterns. The second scenario covers cases when the schema triples $t$ are either above or below $S$. Here, the statistics of $t$ has to be approximated from the data computed for the schema triples $S$ that are included in the statistics. 

The schema triple $t$ is \emph{above} the schema of the statistics $S$ when there is at least one component of $t$, say $c$, such that for all $t'\in\/S:t'\preceq\/t$, $c$ is more general then the components $c'$ that are in the same position in $t'$ as $c$ is in $t$. Note that the component of $t$ referred to as $c$ can be in the role of $s, p$ or $o$. The relationship \emph{below} can be defined similarly to the definition of the relationship \emph{above}. The schema triple $t$ is \emph{below} the schema of the statistics $S$ when there exists at least one component of $t$, say $c$, such that for all $t'\in\/S:t\preceq\/t'$, $c$ is more specific than the components $c'$ that are in the same position in $t'$ as $c$ is in $t$. 

\begin{algorithm}
\normalsize
\begin{algorithmic}[1]
\Function{retrieve-statistics}{$(s,p,o):$ triple-type} $\to$ integer
   \State st $\gets$ 0
   \If{\Call{exists-statistics}{$(s,p,o)$}}
       \State \textbf{return} \Call{get-statistics}{$(s,p,o)$}

   \ElsIf{\Call{above-stat-schema}{$(s,p,o), S$}}
       \State $b_u \gets \{(s_u,p_u,o_u)\ |\ (s_u,p_u,o_u)\in\/S\land\/(s_u,p_u,o_u)\/\preceq(s,p,o)\ \land$ 
       \State \phantom{$b_u \gets \{(s_u,p_u,o_u)\ |\ $}$\not\exists\/(s_u',p_u',o_u')\in\/S\ ((s_u,p_u,o_u)\preceq\/(s_u',p_u',o_u')\ \land$
       \State \phantom{$b_u \gets \{(s_u,p_u,o_u)\ |\ \not\exists\/(s_u',p_u',o_u')\in\/S\ ($}$(s_u',p_u',o_u')\preceq\/(s,p,o))\}$

       \ForAll{$t\in\/b_u$} 
           \State $\text{st} \gets \text{st}+\Call{get-statistics}{(t)}$
       \EndFor

   \ElsIf{\Call{below-stat-schema}{$(s,p,o), S$}}
       \State $b_l \gets \{(s_l,p_l,o_l)\ |\ (s_l,p_l,o_l)\in\/S\land\/(s,p,o)\/\preceq(s_l,p_l,o_l)\ \land$ 
       \State \phantom{$b_l \gets \{(s_l,p_l,o_l)\ |\ $}$\not\exists\/(s_l',p_l',o_l')\in\/S\ ((s_l',p_l',o_l')\preceq\/(s_l,p_l,o_l)\ \land$
       \State \phantom{$b_l \gets \{(s_l,p_l,o_l)\ |\ \not\exists\/(s_l',p_l',o_l')\in\/S\ ($}$(s,p,o)\preceq\/(s_l',p_l',o_l'))\}$
   
       \ForAll{$t\in\/b_l$} 
          \State st $\gets$ \Call{min}{st,\textsc{get-statistics}(t)}
       \EndFor
   \EndIf
   \State \textbf{return} st
\EndFunction
\end{algorithmic}
\caption{Function \textsc{retrieve-statistics}($t:$ triple-type) $\to$ integer}
\label{ret-stat}
\end{algorithm}

The retrieval of the statistical entry for a given schema triple $t=(s,p,o)$ is presented in Algorithm \ref{ret-stat}.  The first part of the algorithm, stated in lines 3-4, retrieves the statistics for the schema triples $t$ that are stored in the statistical index. 

The computation of the statistics, when the schema triple $t$ is above the schema of the statistics, is presented in lines 5-10. First, the set $b_u$ includes schema triples from $S$ that are the closest to $t$, i.e., the schema triples $t_u$ such that $t_u\preceq\/t$ and there is no other $t_u'\in\/S$ which would be between $t_u$ and $t$. In other words, the set $b_u$ includes the upper bounds $t_u$ from $S$ such that $t_u\preceq\/t$. Therefore, $b_u$ contains unrelated schema triples that are the specializations of $t$. We approximate the statistics of a schema triple in lines 9-10 by summing the statistics of the schema triples from $b_u$. We may consequently make mistakes because we count triples that belong to more than one schema triple from $b_u$ multiple times. 

The computation of the statistics when $t$ is below the schema of the statistics $S$ is presented in lines 11-16. It is defined similarly to the case that $t$ is above $S$. The set $b_l$ of the schema triples is computed by selecting the most specific schema triples $t_l\in\/S:t\preceq\/t_l$ that are not interrelated. The statistics of the parameter schema triple $t$ is approximated by using the statistics of the schema triple $t_l\in\/b_l$ with the smallest interpretation. Indeed, the instances of $t$ are also instances of $t_l\in\/b_l$. Therefore, they represent the intersection of the interpretations of all $t_l\in\/S$. Hence, selecting $t_l$ with the smallest interpretation provides an upper bound of the size of the interpretation of $t$. The actual size of $t$ may be smaller. 

\begin{example}
Suppose that we have computed the statistical index for the stored schema of \emph{Simple} by using the procedure \textsc{statistics-stored}. Observe that the same schema would be obtained by using the procedure \textsc{statistics-levels} if the number of levels above and below the stored schema is zero. 

The stored schema comprises the schema triples (person,wasBornIn,location), (person,influences,person) and (philosopher,hasAge,xsd:integer), as well as, the meta-schema including triples (owl:Thing,rdf:type,\-owl:Thing), (owl:Thing,rdfs:sub\-ClassOf,owl:Thing), (owl:Thing,rdfs:subPropertyOf,owl:Thing), etc. However, we will not use the meta-schema in this example. 

The statistical index ST presented in this example solely includes the number of triples for the given schema triples. Recall that the schema triples serve as keys of the statistical index. The complete treatment of the statistics for all types of keys is presented in detail in the following Section \ref{sec-count-keys}.  The entries of the statistical index for the stored schema triples are ST\{(person,wasBornIn,location)\}$=3$, ST\{(person,influences, person)\}$=2$ and ST\{(philosopher,hasAge,xsd:integer)\}$=2$. 

In the case we need the statistics of a schema triple that is included in the index, it can merely be retrieved. For example, the number of instances of (person,wasBornIn, location) is $3$. However, in the case that we need the statistics for a schema triple that is either more specific or more general than the keys of the statistical index, then the statistics are approximated as presented in Algorithm \ref{ret-stat}.  

For example, the schema triple (scientist,wasBornIn,location) is more specific then (person,wasBornIn,\-location), therefore, as described in Algorithm \ref{ret-stat}, (scientist,wasBornIn, location) is below the schema of statistics. In this case, the statistics of (scientist,was\-Born\-In,location) are approximated by the statistics of the least upper bound schema triples from the computed schema of statistics. The schema triple (person,wasBornIn, location) is the only candidate for the closest more general schema triple of (scientist,wasBornIn,location). Hence, the statistics are approximated by using the index entry for (person,wasBornIn,location). $\finbox$ \end{example} 

\section{On counting keys\label{sec-count-keys}}

Let $g\in\/G$ be a knowledge graph stored in a triple-store, and, let $t=(s,p,o)$ be a triple. All algorithms for the computation of the statistics of $g$ presented in the previous section enumerate a set of schema triples $S$ such that for each schema triple $t_s\in\/S$ the interpretation $\llbracket\/t_s\rrbracket^*_g$ includes the triple $t$. For each of the computed schema triples $t_s\in\/S$ the procedure \textsc{update-statistics}$()$ updates the statistics of the key $t_s$ in the statistical index as the consequence of the insertion of the triple $t$ into the graph $g$. 

The triple $t=(s,p,o)$ includes seven keys: $s,p,o,sp,so,po$ and $spo$. These keys are the targets to be queried in the triple-patterns. Let us give an example of a triple-pattern to present the data needed for the estimation of the size of the triple-pattern interpretation. 

\begin{example}
Let $t_p=(?x,$wasBornIn$,$athens$)$ be a triple-pattern including the variable $?x$ in the position of $s$. To compute the number of the triples in $\llbracket\/t_p\rrbracket^*_g$ we first need to know the type of the triple-pattern $t_p$, which can be obtained by retrieving the types of the domain and range of the predicate wasBornIn from the graph $g$. The type of $t_p$ is the schema triple $($person,wasBornIn,location$)$.

The number of triples from $\llbracket\/t_p\rrbracket^*_g$ can be computed by first computing the size of the interpretation $\llbracket\/($person$,$wasBornIn,location$)\rrbracket^*_g$, i.e., the number of \emph{all} instances of $($person,\-wasBornIn,location$)$. Afterward, we need to know the number of different pairs of instances of the predicate wasBornIn and the objects of type location. The type of the key composed of the predicate wasBornIn and an instance of the class location is called \emph{key type}---it is written as $($\underline{person},wasBornIn,location$)$. The keys and the types of keys are defined formally in the sequel. 

The size of $\llbracket(?x,$wasBornIn$,$athens$)\rrbracket^*_g$ can be now estimated as the number of \emph{all possible} triples from $\llbracket\/($person$,$wasBornIn,location$)\rrbracket^*_g$ divided by the number of \emph{different} keys of the type $($\underline{person},wasBornIn,location$)$. In this way we obtain the estimation of the number of triples that have $p$=wasBornIn, $o$=athens, and, arbitrary value of $s$. $\finbox$ 
\end{example}

\subsection{Keys and key types\label{keys-types}}

Let us now define the concepts of the key and the key type. A \emph{key} is a triple $(s_k,p_k,o_k)$ composed of $s_k\in\/I\cup\{\_\}$, $p_k\in\/P\cup\{\_\}$ and $o_k\in\/O\cup\{\_\}$. Note that we use the notation presented in Section \ref{ts-schema}. The symbol ``\_'' denotes a missing component. 

A \emph{key type} is a schema triple that includes the types of the key components as well as the \emph{underlined types} of components that are not parts of keys. However, the underlined types can restrict keys\footnote{See bound/unbound keys in Section \ref{comp-stat-keys}}. Formally, a key type $(s_k,p_k,o_k)$ is a schema triple $(s_t,p_t,o_t)$, such that $s_k\in\llbracket\/s_t\rrbracket^*_g$, $p_k\in\llbracket\/p_t\rrbracket^*_g$ and $o_k\in\llbracket\/o_t\rrbracket^*_g$ for all the components $s_k,p_k$ and $o_k$ that are not ``\_''. The underlined types of the missing components are computed from the stored schema of a knowledge graph. The type inference algorithm is not the subject of the research presented in this paper; more details on the type inference is given \cite{b3s-type-infer}. 

The \emph{complete} type of the key represents the key type where the underlines are omitted, i.e., the complete type of the key is a bare schema triple. 

\begin{example}
The examples of the above defined concepts are presented here.  The triple $($\_,wasBornIn,\-athens$)$ is a key with the components p$=$wasBornIn and o$=$athens, while the component s does not contain a value.  The schema triple $($\underline{person},wasBorn\-In,\-location$)$ is the type of the key $($\_,was\-Born\-In,\-athens$)$. Finally, the complete type of $($\underline{person},was\-Born\-In,location$)$ is the schema triple $($person,wasBornIn,location$)$. $\finbox$ 
\end{example}

\subsection{Computing statistics for keys\label{comp-stat-keys}}

Let us now present the procedure for updating the statistical index for a given triple $t=(s,p,o)$ and the corresponding schema triple $t_t=(t_s,t_p,t_o)$. In order to store the statistics of a triple $t$ of type $t_t$, we split $t_t$ into the seven key types: $(t_s,\underline{t_p,t_o})$, $(\underline{t_s},t_p,\underline{t_o})$, $\ldots$, $(\underline{t_s},t_p,t_o)$, $(t_s,t_p,t_o)$. Furthermore, the triple $t$ is split into the seven keys: $(s,\_,\_)$, $(\_p,\_)$, $\ldots$, $(\_,p,o)$, $(s,p,o)$.  The statistics is updated for each of the selected \emph{key types}. 

There is more than one way of counting the instances of a given \emph{key type}. For the following discussion we select an example of the key type, say $t_k=(\underline{t_s},t_p,t_o)$, and consider all the ways we can count the key $(\_,p,o)$ for a given key type $t_k$. The conclusions that we draw in the following paragraphs are general in the sense that they hold for all key types. 

\begin{enumerate}
\item Firstly, we can either count all keys, including the repeating keys or, we can count only the different keys. We denote these two options by using the descriptive parameters \emph{all} or \emph{distinct}, respectively. 

\item Secondly, we can either count triples of the type $(t_s,t_p,t_o)$, or, the triples of the type $(\top,t_p,t_o)$. In the first case we call counting \emph{bound} and in the second case we call it \emph{unbound}. More detailed description of bound and unbound ways of counting is presented in the following Section  \ref{cnt-bound-unbound}. 
\end{enumerate}

The above stated choices are specified as parameters of the generic procedure \textsc{update-keytype}{(all$|$distinct,bound$|$unbound,$t_k$,$t$)}. The first parameter specifies if we count all or distinct triples.  The second parameter defines the domain of counters, which is either restricted by a given complete type $t_t$ of key type $t_k$, or unrestricted, i.e., the underlined component is not bound by any type. The third parameter is a key type $t_k$, and, finally, the fourth parameter represents the triple $t$. 

Note that we do not need to pass the triple $t$ to the procedure in the case that we count all keys since one is always added to the current statistics of a given key type $t_k$ regardless of the value of $t$.  In the case that we count distinct keys, we need the parameter triple $t$ to extract the key to be inserted in the statistical index for a given key type~$t_k$. 

\subsection{Counting bound/unbound.\label{cnt-bound-unbound}}

The second parameter of the procedure \textsc{update-keytype} determines the possible triples that are taken into account when counting keys of the key type $t_k=(\underline{t_s},t_p,t_o)$. In the case that solely the instances of the type $(t_s,t_p,t_o)$ are used, i.e., the \emph{bound} case, then we do not count instances of $(t_s',t_r,t_o)$ where $t_s'$ is not related to $t_s$. In the case that instances of type $(\top,t_p,t_o)$ are used, that is the \emph{unbound} case, then $s$ in $t=(s,p,o)\in\/g$ can be of any type $\top$. 

\begin{example} 
Let us give an example of the bound and unbound counting. The restriction of the keys by the underlined types of components in the case of bound counting requires each key to be the part of some triple, which is included in the interpretation of the complete type of the key. For example, the key $($\_,wasBornIn,\-athens$)$ of type $($\underline{person},wasBornIn,location$)$ is included in the triple $($plato,wasBornIn,athens$)$ that is an element of the interpretation $\llbracket($person,wasBorn\-In,location$)\rrbracket^*_g$. 

Suppose that the knowledge graph from Figure \ref{simple} contains the triple $($tom,was\-BornIn,paris$)$, where tom is not of the type person but of the type cat. The triple would be taken into account when counting in the unbound way the instances of the key type $($\underline{person},wasBornIn,location$)$. However, the triple $($tom,wasBornIn,paris$)$ is not included in $\llbracket($person,wasBornIn,location$)\rrbracket^*_g$ and is not taken into account when counting in the bound way. $\finbox$
\end{example} 

\subsection{Procedure \textsc{update-statistics}.\label{updt-sttstcs}}

Let us now present the procedure \textsc{update-statistics}($t_t$,$t$) for updating the entry of the statistical index that corresponds to the key schema triple $t_t$ and a triple $t$.  The procedure is presented in Algorithm~\ref{update-statistics}. 

\begin{algorithm}
\normalsize
\begin{algorithmic}[1]
\Procedure{update-statistics}{$t_t$: schema-triple, $t:$ triple}
   \ForAll{key types $t_k$ of $t_t$}
   \State \Call{Update-Keytype}{all,unbound,$t_k$}; 
   \State \Call{Update-Keytype}{all,bound,$t_k$}; 
   \State \Call{Update-Keytype}{dist,unbound,$t_k,t$};
   \State \Call{Update-Keytype}{dist,bound,$t_k,t$};
   \EndFor
\EndProcedure
\end{algorithmic}
\caption{Procedure \textsc{update-statistics}($t_t$:schema-triple, $t:$ triple)}
\label{update-statistics}
\end{algorithm}

The parameters of the procedure \textsc{update-statistics} are the schema triple $t_t$ and the triple $t$ such that $t_t$ is a type of $t$. The \textsc{for} statement in line 1 generates all seven key types of the schema triple $t_t$. The procedure \textsc{update-keytype} is applied to each of the generated key types with the different values of the first and the second parameters. 

We have enumerated the calls of all possible types of procedure \textsc{update-keytype}. However, we expect that the subset of these calls is used for the computation of the statistics of a knowledge graph. 

\section{Experimental evaluation}

In this section, we present the evaluation of the algorithms for the computation of the statistics on the two example knowledge graphs. First of all, we present the experimental environment used for the computation of statistics. Secondly, the statistics of a simple knowledge graph introduced in Figure \ref{simple} is presented in Section \ref{simple-dtset}. Finally, the experiments with the computation of the statistics of the Yago are presented in Section \ref{exp-yago}. 

\subsection{Testbed description}

The algorithms for the computation of the statistics of graphs are implemented in the open-source system for querying and manipulation of graph databases \emph{epsilon} \cite{epsilon}. \emph{epsilon} is a lightweight triple-store management system based on Berkeley DB \cite{oracle-bdb}. It can execute \emph{basic graph-pattern queries} on the triple-store that includes up to 1G triples. 

\emph{epsilon} was primarily used as a tool for browsing ontologies. The operations implemented in \emph{epsilon} are based on the sets of identifiers $I$, as they are defined in Section \ref{tsform-identifiers}. The operations include computing transitive closures based on some relationship (e.g., the relationship rdfs:subClassOf), level-wise computation of transitive closures, and computing least upper bounds of the set elements with respect to the stored ontology. These operations are used in the procedures for the computation of the triple-store statistics.
 
\subsection{Simple knowledge graph\label{simple-dtset}}

The knowledge graph \emph{Simple}, introduced in Figure \ref{simple}, includes 33 triples describing the scientists and the philosophers. We have omitted a large part of Yago meta-schema to provide a simple example for the demonstration of the characteristics of the algorithms for the computation of statistics. The dataset \emph{Simple} is available from the \emph{epsilon} data repository \cite{epsilon-datasets}.

Table  \ref{eval-simple} describes the properties of the schema triples computed by the three algorithms for the computation of the statistics. Each line of the table represents an evaluation of the particular algorithm on \emph{Simple}. The algorithms denoted by the keywords \emph{ST}, \emph{AL} and \emph{LV} refer to the algorithms \textsc{statistics-stored}, \textsc{statistics-all} and \textsc{statistics-levels} presented in the Sections \ref{1st-stat-proc}-\ref{3rd-stat-proc}, respectively. The running time of all the algorithms used in the experiments was below 1ms. 

The second and the third columns are \emph{\#ulevel} and \emph{\#llevel}. They represent the number of levels above and below the schema triples of the stored schema graph that are used in the algorithm \emph{LV}. Hence, these two parameters are not relevant to the algorithms \emph{ST} and \emph{AL}. 

The fourth and fifth columns of the table state the number of schema triples included in the schema graph of the computed statistics. The column name \#bound represents the number of schema triples computed with the \emph{bound} type of counting, and, the column named \#unbound stores the number of schema triples computed by using \emph{unbound} type of counting. 

\begin{table}[h!]
\centering
    \begin{tabular}{ | l | l | l | l | l | p{5cm} |}
    \hline
    algorithm & \#ulevel & \#dlevel & \#bound & \#unbound \\ 
    \hline
    ST & - & - & 63  & 47 \\ \hline
    AL & - & - & 630 & 209 \\ \hline
    LV & 0 & 0 & 63  & 47 \\ \hline
    LV & 0 & 1 & 336 & 142 \\ \hline
    LV & 0 & 2 & 462 & 173 \\ \hline
    LV & 1 & 0 & 147 & 72 \\ \hline
    LV & 1 & 1 & 476 & 175 \\ \hline
    LV & 1 & 2 & 602 & 202 \\ \hline
    LV & 2 & 0 & 161 & 76 \\ \hline
    LV & 2 & 1 & 490 & 178 \\ \hline
    LV & 2 & 2 & 616 & 205 \\ \hline
    \end{tabular}
\caption{The evaluation of the Algorithms 1-3 on the \emph{Simple} triple-store}
\label{eval-simple}
\end{table}

Algorithm \emph{ST} computes the statistics solely for the stored schema graph. The algorithm \ref{statistics-stored} can be compared to the relational approach, where the statistics are computed for each relation. Algorithm \emph{AL} computes the statistics for all possible schema triples. The computed statistics can be used for the precise estimation of the size of the query results. However, as it can be observed from the results of the \emph{Simple}, the number of schema triples computed by this algorithm is significant, even for this small instance. In algorithm \emph{LV}, the number of levels above and below the stored schemata determines the size of the schema graph. The statistics of the schema triples that are above the stored schema triples are usually computed since we are interested in having the global statistics based on the most general schema triples from the schema graph. 

Let us give some comments about the size of the schema graph computed by each particular algorithm used for the computation of the statistics. The schema of \emph{Simple} has only three levels, i.e., the maximal \#ulevel and \#dlevel both equal 2.  Therefore, the number of schema triples of the schema graph does not change if the values \#ulevel and \#dlevel are further increased. 

Note that the algorithm \emph{ST} gives the same results as the algorithm \emph{LV} with the parameters \#ulevel=0 and \#dlevel=0. Indeed, the statistics computed by the algorithm \emph{LV} uses 0 levels above and 0 levels below the schema triples from the stored schema graph, i.e., the stored schema graph itself.  

\subsection{Yago-S knowledge graph\label{exp-yago}}

In this section, we present the evaluation of the algorithms for the computation of the statistics of the Yago-S knowledge graph---we use the core of the Yago 2.0 knowledge base \cite{Hoffart201328}, including approximately 25M (mega) triples. The Yago-S dataset together with the working environment for the computation of the statistics is available from the \emph{epsilon} data repository \cite{epsilon-datasets}. Let us now describe some characteristics of the Yago knowledge graph, and a set of test evaluations of the algorithms for the computation of the statistics of the Yago-S knowledge graph. 

Yago includes three types of classes: the Wordnet classes, the Yago classes, and the Wikipedia classes. There are approximately 68000 Wordnet classes used in Yago that represent the top hierarchy of the Yago taxonomy. The classes that are defined within the Yago dataset are used mostly to link the parts of the datasets. For example, they link the Wikipedia classes to the Wordnet hierarchy. There are less than 30 newly defined Yago classes. The Wikipedia classes are defined to represent group entities, the individual entities, or some property of the entities. There are about 500000 Wikipedia classes in Yago. 

We use solely the Wordnet taxonomy and newly defined Yago classes for the computation of the statistics. We do not use the Wikipedia classes since, in most cases, they are very specific. The height of the taxonomy is 18 levels, but there are a small number of branches that are higher than 10. While there are approximately 571000 classes in Yago, it includes only a small number of predicates. There are 133 predicates defined in Yago 2.0 \cite{Hoffart201328}, and there are only a few sub-predicates defined; therefore, the structure of the predicates is almost flat. 

Table \ref{eval-yago} presents the execution of the algorithms \textsc{statistics-stored}, \textsc{statistics-all} and \textsc{statistics-levels}, which we denote as ST, AL and LV, respectively. Each line represents two executions of a given algorithm with the specified parameters: in the first execution, we use \emph{bound} type of counting, and, in the second execution, we use \emph{unbound} way of counting. 

The first column states the algorithm used in the two executions presented in the given line of the table. The column named \#bound stores the number of the schema triples of the statistics computed for the bound mode of counting. The column named time-b represents the time in hours used for the computation of the statistics. Similarly, the column named \#unbound includes the number of the schema triples of the statistics computed in the unbound mode, and the column named time-u presents the number of hours that were needed for the computation of the statistics. Finally, the columns \#ulevel and \#dlevel specify the number of levels above and below the schema triples from the stored schema graph that are used in the algorithm LV. In the following paragraphs, we give some general comments about the behavior of algorithms in experiments. 

\begin{table}[h!]
\centering
    \begin{tabular}{ | l | l | l | r | r | r | r | }
    \hline
    algorithm & \#ulevel & \#dlevel & \#bound & time-b & \#unbound & time-u \\ 
    \hline
    ST & - & - & 532 & 1 & 405 & 1 \\ \hline
    AL & - & - & >1M & >24  & >1M & >24 \\ \hline
    LV & 0 & 0 & 532 & 4.9 & 405 & 4.9 \\ \hline
    LV & 1 & 0 & 3325 & 6.1 & 1257 & 5.4 \\ \hline
    LV & 2 & 0 & 8673 & 8.1 & 2528 & 6.2 \\ \hline
    LV & 3 & 0 & 12873 & 9.1 & 3400 & 6.8 \\ \hline
    LV & 4 & 0 & 15988 & 10.9 & 3998 & 7.1 \\ \hline
    LV & 5 & 0 & 18669 & 11.7 & 4464 & 7.5 \\ \hline
    LV & 6 & 0 & 20762 & 12.5 & 4819 & 7.8 \\ \hline
    LV & 7 & 0 & 21525 & 13.1 & 4939 & 7.9 \\ \hline
    LV & 0 & 1 & 116158 & 5.9 & 27504 & 5.4 \\ \hline
    LV & 1 & 1 & 148190 & 7.8 & 34196 & 6.3 \\ \hline
    LV & 2 & 1 & 183596 & 10.5 & 40927 & 7.4 \\ \hline
    LV & 3 & 1 & 207564 & 12.7 & 45451 & 7.8 \\ \hline
    LV & 4 & 1 & 225540 & 14.4 & 48545 & 8.3 \\ \hline
    LV & 5 & 1 & 239463 & 15.6 & 50677 & 8.7 \\ \hline
    LV & 6 & 1 & 250670 & 17.3 & 52376 & 9.1 \\ \hline
    LV & 7 & 1 & 255906 & 19.8 & 53147 & 9.2 \\ \hline
    LV & 0 & 2 & 969542 & 7.6 & 220441 & 6.6 \\ \hline
    LV & 1 & 2 & >1M & >24 & >1M & >24 \\ \hline
    \end{tabular}
\caption{The evaluation of the Algorithms 1-3 on the Yago knowledge graph}
\label{eval-yago}
\end{table}

Let $t=(s,p,o)$ be a triple for which we would like to compute the statistics. The computation of the schema triples that represent the types of $t$ depends on 1) the algorithm employed, and 2) on the enrollment of a triple $t$ in the triple store schemata. The extent of a schema graph taken into account for the computation of the statistics depends on the concrete algorithm and the values of its parameters. For instance, the extent of selected schema triples in the algorithm \textsc{statistics-levels} depends on the number of selected levels below and above the stored schema. The enrollment of a given triple $t$ in the knowledge graph schema is defined by the relationships among the $s$, $p$ and $o$ components of $t$ to their types, expressed using the predicate rdf:type. If $t$'s components $s$, $p$, or $o$ have many types specified in the triple store, then $t$ has a large set of the possible types. 

Therefore, the computation time for updating the statistics for a given triple $t$ depends on the selected algorithm and the complexity of the triple enrollment into the conceptual schemata. In general, the more schema triples that include $t$ in their interpretation, the longer the computation time is. For example, the computation time increases significantly in the case that lower levels of the ontology are used in the computation of the statistics, simply because there is a large number of schema triples (approx.\ 450K) on the lower levels of the ontology. Finally, the computation of the statistics for the triples that describe people and their activities takes much more time than the computation of the statistics for the triples that represent links between web sites, since the triples describing people have richer relationships to the schema than the triples describing links among URIs. 

Let us now give some comments on the results presented in Table \ref{eval-yago}. The algorithm ST for the computation of statistics of a knowledge graph, including 25 Mega triples, takes about 1 hour to complete in the case of bound type and unbound type of counting. The computation of the algorithm AL does not complete because it consumes more than 32 GB of main memory after the schema triples of 2 Mega triples are computed and their statistics updated. 

The algorithm \textsc{statistics-levels} can regulate the amount of the schema triples that serve as the framework for the computation of the statistics. The number of the generated schema triples increases when more levels either above or below the stored schemata are taken into account. We executed the algorithm with the parameter \#dlevel$=0,1,2$. For each fixed value of \#dlevel, the second parameter \#ulevel value varied from $0$ to $7$. The number of generated schema triples increases by about a  factor of 10 when the value of \#dlevel changes from $0$ to $1$ and from $1$ to $2$. Note also that varying \#ulevel from $0$ to $7$ for each particular \#dlevel results in an almost linear increase of the generated schema triples. This is because the number of schema triples is falling fast as they are closer to the most general schema triples from the knowledge graph, i.e., when the value \#ulevel increases. 

\section{Related work\label{related-work}}

A knowledge-based system is composed of two parts: a knowledge base and an inference engine \cite{Brachman2004KnowledgeRA}. A knowledge base (abbr.\ KB) includes the knowledge part and the data part. The knowledge part of a KB includes the definitions of concepts in the form of an ontology, the terminological axioms that state the equivalences and subsumptions among the defined concepts, and rules that involve defined concepts. The data part of KB includes the facts often called assertions. An inference engine of a knowledge-based system can have various forms. The most common inference engine, developed for rule-based systems, uses deductive reasoners based on backward and forward rule chaining to derive interesting consequences \cite{Lenat1995}. The inference engine in systems based on the description logic uses the satisfiability and subsumption to model reasoning \cite{Baader2002}. This kind of automated reasoning system is often called a \emph{classifier}.   
 
From the perspective of a knowledge-based system, a knowledge graph is a knowledge base that uses the graph as the data structure for the representation of the data and knowledge. As already stated in the introduction section, we suppose that a knowledge graph is defined by using the data representation languages RDF \cite{rdf}, RDF-Schema \cite{rdfschema}, and OWL \cite{owl}. We expect that the knowledge graph is stored in a triple-store system \cite{Neumann:2010:RES:1731351.1731354,Gurajada:2014:TDS:2588555.2610511,jena-tdb,Harth:2007,4store,virtuoso,Zeng:2013}. A triple-store system used for storing and querying a knowledge graph is a knowledge-based system. A querying system can be used, to some extent, as a tool for browsing and inferring the new knowledge from a triple-store. Moreover, recent triple-stores offer some capabilities to infer new knowledge from the knowledge graph that includes RDF-Schema statements \cite{Kaoudi2015}. Finally, we expect that triple-store systems will gradually incorporate the inference engines such as a deductive reasoner and a classifier. 

The existing knowledge-based systems use indexes similar to those from the database management systems (abbr.\ DBMS) to speed up the access to assertions \cite{cyc}, the matching of rules \cite{Forgy1982}, and practically in any situation where large quantities of similar data entities need to be accessed \cite{Brachman2004KnowledgeRA}. However, we do not know of any use of the statistics in the implementation of inference engines of knowledge-based systems. On the other hand, the statistics are used in practically any DBMS, including the triple-store systems, to provide the means for cost-based query optimization. Let us now present the use of statistics in relational database systems and triple-store systems.

In the System R, the statistics were used for the estimation of the selectivity of simple predicates and join queries in query optimization process \cite{Selinger:Etal:1979:SIGMOD}. The gathered statistics include for each relation the cardinality, the number of pages, and the fraction of the pages that store tuples. Further, for each index System R stores the number of distinct keys and the number of pages of the index. The model of the statistics used in System R presupposes the uniform distribution of the attribute values. In the case that the attribute values are not distributed uniformly, the statistics based on the above parameters can give inaccurate selectivity estimations \cite{Shapiro:Etal:1984:SIGMOD}. 

The non-uniform distribution of the attribute values can be captured better by using the equidistant histograms \cite{Christodoulakis:1983:SIGMOD}. The set of attribute values are split into equidistant intervals, and the number of attribute values in each interval is stored. Piatetsky-Shapiro and Connell show in \cite{Shapiro:Etal:1984:SIGMOD} that the equidistant histograms fail very often to give a precise estimation of the selectivity of simple predicates. As a better alternative, they propose the use of the histograms that have equal heights, instead of equal intervals. Furthermore, they show that the precision of the selectivity estimation can be easily improved by increasing the number of intervals in the histogram. 

Our schema-based approach to the computation of the knowledge graph statistics currently uses the counters for all and distinct triples of each particular schema triple from the schema graph. However, the histograms can be used in a similar way they are used in the relational systems. For each particular key type of a given schema triple, we can construct a histogram.

Triple-stores were initially designed as schema-less databases for the representation of graphs. However, the RDF data model \cite{rdf} was extended with the RDF-Schema \cite{rdfschema} that allows for the representation of knowledge bases. Consequently, RDF graphs can separate the conceptual and instance levels of the representation, i.e., between the TBox and ABox \cite{Baader2002}. Moreover, RDF-Schema can serve as the means for the definition of taxonomies of classes (or concepts) and predicates, i.e., the roles of a knowledge representation language. Let us now present the existent approaches to collect the statistics of triple-stores. 

Virtuoso \cite{virtuoso} is based on relational technology. The relational statistics are in Virtuoso computed periodically; however, when processing SPARQL queries, the statistics are computed in real-time by inspecting one of 6 indexes \cite{virtuoso-rdf-1}.  The size of conjunctive queries that can include one comparison operation ($\ge$, $>$, $<$, or $\le$) is estimated by counting the pointers of index blocks satisfying the conditions on the complete path from the root to the leaf block of the index. In the case there are no conditions in a query, then sampling ($1\%$ of triples) is used to estimate the size of the query---we choose pointers of the index blocks at random. The results of query estimation are always stored in the index so that they are available for the following requests.

RDF-3X \cite{Neumann:2010:RES:1731351.1731354} is a centralized triple-store system. The triples are first converted to numeric ids and then stored directly in B+ trees.  RDF-3X uses six indexes for each ordering of triple-store columns S, P, and O. B+ tree indexes are customized in the following ways. Firstly, triples are stored directly in the leaves of B+ trees. Secondly, each index uses the lexicographic ordering of triples, which provides the opportunity to compress triples in leaves by storing only the differences between the triples. RDF-3X includes additional aggregated indexes where the number of triples is stored for each particular instance (value) of the prefix for each of the six indexes. Aggregate indexes can be used for the selectivity estimation of arbitrary triple patterns. They are converted into selectivity histograms that can be stored in the main memory to improve the performance of selectivity estimation. Furthermore, to provide a more precise estimation of the size of queries in the presence of correlated predicates, frequent paths are determined, and their cardinality is computed and stored. 

Example \ref{example1} suggests that the use of semantic information can improve the selectivity estimation of joins in comparison to the selectivity estimation based on the aggregate SPO indexes from RDF-3X \cite{Neumann:2010:RES:1731351.1731354}. The correlation between two triple-patterns used in a join can be captured by inspecting the types of triple-patterns. In the case that the type of one triple-pattern is specialized because of the type of the second triple-pattern, we can use the statistics for the specialized type to improve the join selectivity estimation. However, RDF-3X can capture the correlations between joining triple-patterns by the use of the statistics precomputed for frequent paths. 

TriAD \cite{Gurajada:2014:TDS:2588555.2610511} is a distributed triple-store implemented on shared-nothing servers running centralized RDF-3X \cite{Neumann:2010:RES:1731351.1731354}. The triple-store is partitioned utilizing a multilevel graph partitioning algorithm \cite{Karypis:1999} that generates graph summarizations. These are further used for query optimization as well as during the query execution to enable join-ahead pruning. Six distributed indexes are generated for each of the SPO permutations. Partitions are stored and indexed on slave servers while the summary graph is stored and also indexed at the master server. The statistics of the triple-store is stored locally for the local partitions, and, globally, for the summary graph. In both cases, the cardinality is stored for each value of S, P, and O, and, for the pairs of values SP, SO and PO. Furthermore, the selectivity of joins between P$_1$ and P$_2$ predicates is also stored in the distributed index on all the slave servers and for the summary graph on the master server. 

The optimizer of Jena \cite{Stocker:Etal:2008:WWW08} was the first to use the statistics based on some form of semantic data. The statistics are used for the computation of the selectivity estimations in the query optimization algorithm that chooses the ordering of joins for basic graph-patterns. The proposed greedy method represents a balance between the efficiency of the query execution plan and the computation cost of query optimization. The statistics that are used for the heuristics do not cover all types of queries but represent the most commonly used types of queries. For the triple-patterns, the sizes of the main types of bound components and the number of different values of bound components are computed. For instance, the exact statistics are computed for each value of a bound predicate, i.e., the number of triples, including the particular predicate, is computed for each predicate. For the computation of joins, the statistics are computed for each pair of the \emph{related} predicates. Similarly to our approach, the domains and the ranges of predicates are used to select related pairs of predicates that can actually appear in joins. As in the case of the triple-patterns, the sizes of join queries, i.e., the sizes of joined triple-patterns, is computed for each possible type of join. 

The work of Neumann and Moetkotte on characteristic sets \cite{c-sets} (abbr. CS) can be seen as the extension of the work on the query optimizer of the Jena environment. A characteristic set captures the common properties of a set of objects. For a given subject $s$ the CS includes the predicates $\{p|(s,p,o)\in\/TS\}$ where $TS$ stands for triple-store. There is only a small number of CS-s for a given triple-store. The statistics of a triple-store TS are gathered for the CS-s of TS. The CS-s of data and the CS-s of queries are computed. A given star-shaped SPARQL query retrieves instances of all the CS-s that are the super-sets of a CS of a query. In general, the queries can be decomposed into the star-shaped sub-queries. Besides the statistics of CS-s, the number of triples with a given predicate is computed for each of the CS and each predicate. The size of joins in star-queries can be accurately estimated in this way. 

Gubichev and Neumann further extended the work on characteristic sets (abbr. CS) in \cite{gubichev14}. CS-s are organized in a hierarchy. The first level includes all CS-s of a given triple-store. The next level includes the cheapest CS-s that are the subsets of CS-s from the previous level. The selectivity estimation is defined using the statistics of characteristic sets. The hierarchical characterization of CS-s allows precise estimation of joins in star queries. CS of a star query has references to the cheapest CS-s (subsets) that include predicates of the joined triples. The join ordering algorithm selects the cheapest CS-s on the next level of the CS hierarchy in each iteration step, and, in this way, determines the next join in the constructed left-deep query plan. Therefore, the algorithm complexity is linear. General SPARQL queries are handled by first splitting the queries into star sub-queries that are related by a form of a foreign key. Furthermore, the characteristic pairs of CS-s are identified, and the number of instances (statistics) is stored for each characteristic pair. The abstracted form of query graph where meta-nodes replace star queries is optimized by using the regular dynamic programming algorithm together with the statistics for characteristic pairs. 

The computation of the statistics of a triple-store by using the characteristic sets can be seen as a reverse-engineering approach that unravels the structure of a triple-store. The obtained structure of a triple-store should be closely related to the schema graph that defines the semantic structure of the triple-store. Furthermore, the hierarchical characterization of CS-s unravels the classification hierarchy of classes but also the specialization hierarchy of the schema triples. Therefore, our statistics based on the schema triples from several levels around the stored schema triples can handle the correlations that appear in the star queries similarly as proposed by Gubichev and Neumann \cite{gubichev14}.

\section{Conclusions}

This paper presents a new method for the computation of the statistics of knowledge graphs. The statistics are based on the schema graph, the size of which can be tuned to obtain either cruder or more precise statistics. The smallest schema graph that can be used as the framework for the computation of the statistics is the stored schema graph, while the largest schema graph corresponds to all possible schema triples from a given knowledge graph. In between, we can set the number of levels above and below the schema triples from the stored schema graph to be included.

The computation of the statistics of the core of Yago, including around 25M triples, takes from 1-24 hours depending on the size of the resulted schema graph of the statistics. The experiments were run on a low-cost server with 3.3GHz Intel Core CPU, 16GB of RAM, and 7200 RPM 1TB disk. Therefore, it is feasible to compute statistics even for large knowledge graphs. Note that knowledge graphs are not frequently updated so that the statistics can be computed in a batch job. 

The current implementation of algorithms for the computation of the statistics of knowledge graphs is not tuned for performance. While it includes some optimizations, for instance, the transitive closures of specific sets of classes are cached in a main memory index, there is a list of additional tuning options at the implementation level that will speed-up the computation of the statistics in practice.

The statistics of a knowledge graph are primarily used for the estimation of the results of the SPARQL queries. However, the selectivity estimation of the graph algebra operations was not the focus of the research presented in this paper. We have developed the type-checking procedure to determine the types of query expressions, and we performed a set of experiments with the estimations of the size of the graph algebra expressions. The initial results show that the accuracy of the estimations can be improved by computing more precise statistics of the knowledge-graph. 

Finally, we use the statistics of knowledge graphs in the algorithms for the distribution of the triple-store to an array of servers. The \emph{skeleton graph} of the knowledge graph is computed to include the ``skeleton'' schema triples, i.e., the schema triples that have the interpretations of the appropriate size to be used as the units of the distribution. Therefore, we obtain a schema graph where all the schema triples have a similar number of instances. The distribution of the triple-store is then based on the partitioning of the skeleton graph. 

\section{Acknowledgments}

The authors acknowledge the financial support from the Slovenian Research Agency (research core funding No. P1-00383).

\bibliographystyle{abbrv}
\bibliography{sm}

\end{document}